\documentstyle[12pt]{l-aa}

\def\fe{$[Fe/H]$}
\def\arcsec{$\,^{\prime\prime}$}
\def\arcmin{${\,^\prime}$}
\begin{document}
   \thesaurus{06         
              (08.05.3;  
               08.06.3;  
               08.08.2;  
               08.22.3)} 
   \title{A photometric study of the field RR Lyrae stars \\
                 AW Dra, AQ Lyr and CN Lyr}

   \author{V. Castellani 
           { \inst{1,}\inst{2,}\inst{3}}
 \and  A. Di Paolantonio
          \inst{2}
 \and  A. M. Piersimoni 
	  \inst{2}	 
 \and  V. Ripepi
          {\inst{1,}\inst{4}}
          }
          
   \offprints{V. Ripepi}

   \institute{Dipartimento di Fisica, Universit\`a di Pisa, Piazza Torricelli 2, I-56100 Pisa, Italy
    \and Osservatorio Astronomico di Collurania, Via M. Maggini, I-64100 Teramo, Italy  
    \and Istituto Nazionale di Fisica Nucleare, LNGS, 67010 Assergi, L'Aquila, Italy
    \and Osservatorio Astronomico di Capodimonte, Via Moiariello 16, I-80131 Napoli, Italy
}


   \maketitle

   \begin{abstract}

We present CCD  lightcurves  for the three field 
RR Lyrae AQ Lyr, CN Lyr and AW Dra observed at the Teramo TNT
telescope. Stellar temperatures have been derived from the
observed mean colors, allowing a comparison
with recent theoretical predictions on RR Lyrae lightcurves. As for the two metal rich
pulsators in the sample we suggest that CN Lyr has been possibly
misclassified, being a c-type pulsator.
This gives a warning against the possible contamination with
first overtone (FO) pulsators of observational samples of field ab type stars.
On the other hand, if CN Lyr belongs to the low amplitude,
b-type fundamental pulsators, it shows a lightcurve with a slow
risetime not jet predicted by theory.
AQ Lyrae appears as a metal rich ab type pulsator, with the 
expected luminosity but with amplitude lower than predicted by theory.
Finally, we find that AW Dra  behaves as a metal poor fundamental (F) pulsator, 
crossing the instability strip well above the corresponding Zero Age Horizontal 
Branch (ZAHB) luminosity level.

   \end{abstract}

%
%

\section{Introduction}

RR Lyrae radial pulsators have been early recognized 
as interesting stellar objects, widely used as population tracers and 
distance indicators. In particular, the pulsational behavior
of metal poor RR Lyrae in galactic globular clusters has been a
most debated argument in the literature, stimulating a large 
amount of investigations both on observational and theoretical grounds.
However, since the pioneering paper by Preston (1959), one knows 
that field RR Lyrae in the solar neighborhood have a metal
rich component, not observed in galactic globulars, which appears
characterized by objects with peculiarly short periods.  
The origin of such a behavior has been recently investigated by Bono 
et al. (1997a, BCCIM hereafter),  who presented detailed theoretical 
predictions on the shape of lightcurves for different assumptions
about star masses, chemical compositions, luminosities and temperatures. 
Accordingly, one can foresee the possibility of connecting several
features of the lightcurves (as the rising time and the occurrence 
of bumps/dips) to the structural parameters of the pulsating stars,
testing the reliability of theoretical predictions and, possibly,
deriving information on the evolutionary status of the observed objects.

However, one finds that several lightcurves 
available in the literature for metal rich, field RR Lyrae,
rely on old photographic photometry, whose intrinsic inaccuracy
does not allow a detailed analysis of the predicted features.
Taking advantage of the CCD technology, one may now
obtain rather accurate photometry  even in connection with small telescopes.
In this paper we report the first results of a similar program started
at the Teramo Astronomical Observatory with the use of the 72 cm
Teramo/Scuola Normale Telescope (TNT).

\begin{figure*}
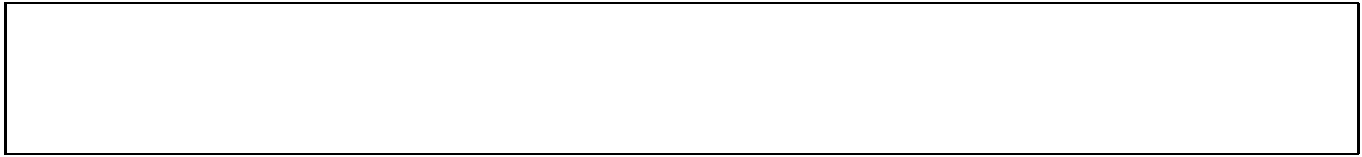

\picplace{2cm}
\caption[]{Identification maps for AW Dra, AQ Lyr and CN Lyr; comparison stars 
are also indicated}
\label{fig1}
\end{figure*}

From the General Catalog of Variable Stars ( 4th Ed: Kholopov e al. 1988, 
GCVS4 hereafter) we selected the sample of RR Lyrae more luminous
than $ B \simeq 14$ mag and with known periods which can be observed 
from Teramo
and for which no accurate lightcurves  are available in the literature.
Among these objects, we chose as first targets the two metal rich 
ab-type RR Lyrae (RRab)
AQ Lyr and CN Lyr. For AQ Lyr one has $\Delta S=1.15$ i.e. 
$\fe=-0.59\pm0.08$ (Z=0.006) (Suntzeff et al. 1994), 
while for CN Lyr, Layden et al. (1996) provide $\fe=-0.26\pm0.07$ (Z=0.01). 
To our knowledge, for AQ Lyr one finds in the literature 
only seven photographic BV measurements (Sturch 1966), whereas CN Lyr
has photoelectric, but rather scattered,  UBV lightcurves 
by Oosterhoff (1962). To these two target stars we added 
the RRab AW Dra for which neither a lightcurve  nor a metallicity
estimate were encountered in the literature.

\begin{table}
\picplace{2cm}
\caption[]{
\small
$B$ and $V$ magnitudes for comparison stars.
\label{tab1}
}
\label{tab1}
\end{table}

\section{Observations and data reductions}

The TNT telescope is equipped with a Tektronics CCD 512x512 pixels, with a total
field of view of 4 x 4\arcmin. Observational data  were 
secured mainly during the 1995 campaigns:  AQ Lyr from July 22 to 27; 
CN Lyr from July 25 to 27  (plus June 10, 1996); AW Dra from July 
19 to 25.  Single exposure 
times ranged from 30-90 sec in  V to 120-300 in B filter,
with a  typical seeing  of about 2.5 \arcsec. 
Several Bias and flat-field frames were taken at the beginning and at 
the end of each night; following the usual procedures, these frames were 
used to pre-reduce observational data.

In order to minimize the observation time and, consequently, to increase the
sampling of the lightcurves, the magnitude of each variable has been
obtained by evaluating the  difference in magnitude between the target star 
and suitable comparison stars in the field of view of the telescope. 
Final data in standard B,V magnitudes have been finally obtained calibrating 
the comparison stars with   several 
Landolt (1992) stars on three nights with good photometric conditions.
Note that in this way one is also speeding up the reduction procedure, 
since no differential atmospheric extinction has to be taken into account
in the magnitude differences between target and comparison stars.

\begin{figure}
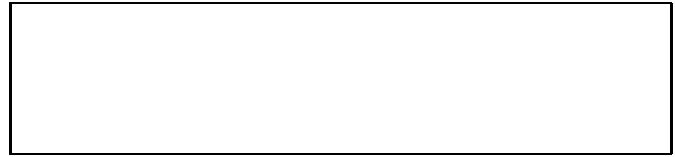

\picplace{2cm}
\caption[]{Magnitude difference in the $B$ band between 
comparison stars C2 and C1 of AW Dra; see the text for a short discussion.}
\label{fig1bis}
\end{figure}

The magnitudes of  stars in each frame have been
obtained through a suitable procedure for the ESO-MIDAS reduction package.
Figure ~\ref{fig1} gives the identification maps for the three variables 
and for the chosen comparison stars. As shown in the figure, in all 
fields we chose two comparison stars. However, the comparison 
star C2 of AW Dra is likely to be a variable star (see Fig.~\ref{fig1bis}), 
whereas the non-variability of star C1 is ensured by the constancy 
of the AW Dra minima; so only the comparison star C1 was used.

In the two other cases, we derived a lightcurve 
with respect to each of the two comparison stars, thus averaging
the two lightcurves to produce the final result. 
We found that differences between the two averaged curves are
of the order of 0.01-0.02 magnitude.
The magnitudes of the comparison stars in Fig.~\ref{fig1} 
are given in Table~\ref{tab1}.  

\section{Lightcurves, amplitudes, mean magnitudes and colors}

\begin{figure}
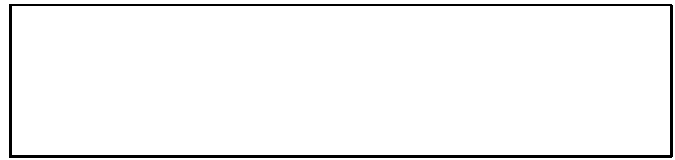

\picplace{2cm}
\caption[]{B and V lightcurves for AW Dra, AQ Lyr, CN Lyr}
\label{fig2}
\end{figure}

By using periods given by GCVS4 one finally finds the 
lightcurves for AW Dra, AQ Lyr and CN Lyr  reported in 
Fig.~\ref{fig2}. Selected photometric data 
are reported in Table~\ref{tab2},~\ref{tab3},~\ref{tab4}.

\begin{table*}
\picplace{2cm}
\caption{
\small
$B$ and $V$ magnitudes for AW Dra.
\label{tab2}
}
\end{table*}

\begin{table*}
\picplace{2cm}
\caption{
\small
$B$ and $V$ magnitudes for AQ Lyr.
\label{tab3}
}
\end{table*}

\begin{table*}
\picplace{2cm}
\caption{
\small
$B$ and $V$ magnitudes for CN Lyr.
\label{tab4}
}
\end{table*}

\begin{figure}[h]
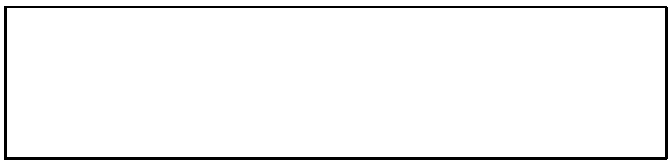

\picplace{2cm}
\caption[]{Comparison with previous photometry (full squares) for AQ Lyr (top) and 
CN Lyr (bottom). For both variables new data are reported as open circles.}
\label{fig3}
\end{figure}

For each variable, the lightcurves in $B$ and $V$ were fitted by means 
of smoothing splines, shown as a solid line in the quoted figures. 
From these fits we finally derived 
for the three RR Lyrae stars the amplitudes in $V$ and $B$ and the mean 
magnitudes and colors reported in Table~\ref{tab5}. 
Figure ~\ref{fig3} (top) compares the present lightcurve for 
AQ Lyr with previous measurements given by Sturch (1966), while 
Fig.~\ref{fig3} (bottom) right compares  the lightcurve 
CN Lyr with the photometry given by Oosterhoff (1962). 
One finds a reasonable agreement together with
the evidence for the great improvement afforded by the use of CCD
even in rather small telescopes as the one used in this investigation.

As a final point, one may compare present results with the rich sample of
lightcurves presented by Lub (1977) for field stars. For each
of our 3 variables one finds in Lub's sample stars with similar
periods and with quite similar lightcurves, supporting the reality
of the various features and showing that in all cases we are
dealing with "typical" field RR Lyrae pulsators.  
\section{Comparison with the theory of pulsation}

In recent times the use of  non-linear, non-local, time-dependent 
convective models by Bono and coworkers has produced a rather large 
amount of theoretical predictions concerning RR Lyrae pulsations.
(See the most  recent papers by Bono et al. 1997a (BCCIM), 1997b (BCCM) 
and reference therein). 
Taking advantage of such a theoretical framework, 
in this section we will discuss the pulsational properties of the
three studied variables.

\begin{table*}
\picplace{2cm}
\caption{
\small 
Observational results for AW Dra, AQ Lyr and CN Lyr.
\label{tab5}
}
\end{table*}

\begin{table*}
\picplace{2cm}
\caption{
\small
Selected fundamental quantities for AW Dra, AQ Lyr and CN Lyr.
\label{tab6}
}
\end{table*}

As a first step, Fig.~\ref{fig4} shows the location in the Bailey 
(period-amplitude) diagram for our three variables, as compared 
with selected samples of field RRab and with theoretical 
results from BCCIM and for the selected, labeled choices about the metallicities
 of field stars. Periods and amplitudes of field RR Lyrae are from Blanco (1992), 
while metallicities are from Layden (1994, 1995) and Layden et al. (1996). 
In all three panels our variables are reported with four-ended stars:
going toward lower periods one finds in the order AW Dra, CN Lyr and AQ Lyr. 
One finds that AW Dra is placed in a region where only  
low metallicity pulsators occur. Accordingly one can predict for this
variable a metallicity as low as $\fe \leq$ -1.4. On the contrary, both AQ Lyr and
CN Lyr clearly are members of the high metallicity group, as expected
in particular for CN Lyr ($\fe \sim$ -0.26). One may notice that
both stars lie on the lower envelope of the observed distributions,
well below theoretical expectations even for "young" massive pulsators 
(see Fig. 16b in BCCIM and the discussion in that paper). 
We conclude that, if these stars are "bona fide" ab-type pulsators,
theory has to be improved to account for such kind of unpredicted 
variables.

\begin{figure}[h]
\picplace{2cm}
\caption[]{The Bailey diagram for the three RR Lyrae of this study, in 
comparison with selected theoretical and observational results. For details 
see the text and Fig.s 16a and 16b in BCCIM}
\label{fig4}
\end{figure}

To allow a closer comparison with predicted lightcurves  one can
estimate star temperatures from
the observed colors, provided that the reddenings are known.  
For AQ Lyr Burstein \& Heiles (1978) provided $E(B-V)$ = 0.127 mag.
Since AQ Lyr has $\Delta S $ values by Suntzeff et al. (1994), one may 
use the method by Sturch (1966), as improved 
by Blanco (1992), to test this reddening with an independent estimate. 
As a result we find for AQ Lyr $E(B-V)$ = 0.13 mag, in excellent
agreement with the previous value.
For CN Lyr and AW Dra the literature gives no indications.
To get the missing values, we again used the reddening maps by Burstein \& Heiles 
(1982), obtaining for AW Dra $E(B-V)$ = 0.06 mag and for CN Lyr 
$E(B-V)$ = 0.21 mag. 
According to the quoted authors, the error on these estimates
is of the order of 0.03 mag. 


An alternative way to derive information about reddenings is that of 
using statistical relations as those provided by Caputo \& De Santis (1992).
These authors took advantage of the Lub (1977) sample of field ab type 
variables to derive relations between periods, B amplitudes, metallicities 
and mean dereddened colors.
Using their Eq.(10) we obtain for CN Lyr $(B-V)_0$ = 0.38 mag and thus 
$E(B-V)$ = 0.20 mag in good agreement with the value given by Burstein \& Heiles (1982).
The same procedure for AQ Lyr provides $(B-V)_0$=0.33 mag that 
means $E(B-V)$ = 0.10 mag, slightly lower (about 0.03) than the value 
previously determined 
from Blanco's reddening law and from Burstein \& Heiles (1978) maps but within 
the errors.  
Hence we adopt those estimates in order to compute the temperature 
of the variable in table 6.
For AW Dra there is no available metallicity evaluation in the literature, but we
can estimate lower and upper limits for it from our Fig.~\ref{fig4}.
We get, for $\fe = -1.4$, $(B-V)_0$=0.33 mag and, for $\fe = -1.9$,
$(B-V)_0$=0.32 mag that in terms of reddening means $0.04<E(B-V)<0.05$ mag,
well within the error of  the reddening as derived 
from Burstein \& Heiles (1982) maps.

\begin{table*}
\picplace{2cm}
\parbox{14cm}{
\caption{
\small
Luminosities coming from period-temperature-mass-luminosity relations (see text) by 
BCCM, for the three RR Lyrae investigated in this study; F and FO mean fundamental and first overtone pulsation mode respectively. 
\label{tab7}
}}
\end{table*}

After correction for reddening, mean colors have been evaluated in 
three ways: as intensity-weighted ($<B-V>$ or $<B>-<V>$) and as 
magnitude-weighted 
$(B-V)$; all these values are reported in Table~\ref{tab5}. 
However Bono et al. (1995, BCS hereafter) have once again shown
that the color of the static model does not match exactly any observed mean color.
 To all these mean colors we thus applied the corresponding correction
as tabulated by BCS for Z=0.001; 
we estimated that errors due to the
different metallicity should not exceed few thousandths of magnitude,
thus preserving the general trend 
 $[(B-V)] > [<B-V>] > [(B-V)_{static}] >  [<B>-<V>]$. 
After the three substantially different $(B-V)$ 
values 
for each star become very similar, so we proceed in averaging them and 
we took this mean value as our best estimate of the RR Lyrae mean colors, 
as given by $(B-V)_{mc}$ in Table~\ref{tab5}.
RR Lyrae mean colors have been finally translated in temperatures 
by estimating gravities from the period - gravity relation 
obtained from the  period - temperature - luminosity - mass relation by BCCM
 and using  Kurucz (1992) models. 
One finds $\log g = 3.0$ for both  
AQ Lyr and CN Lyr, whereas for AW Dra one has  $\log g = 2.75 $.
The estimated error on the temperature is about $\pm 100 K$, 
largely dominated by the error on the reddening.
Temperatures for the three RR Lyrae studied in this paper are reported 
in Table~\ref{tab6} together with further selected quantities.

Given period and temperature, one can obtain an estimate of the
star luminosity from the period - temperature - luminosity - mass
relation, provided that suitable assumption on the pulsator masses
are made. By looking into evolutionary constraints one can safely
assume M= 0.53-0.58 $\rm M_{\odot}$  for the two metal rich RR Lyrae, 
and M=0.65-0.75
$\rm M_{\odot}$ for AW Dra. Making use of the relations given by BCCM 
(corrected by $\rm \Delta \log P \sim +0.02$ for fundamental 
metal rich pulsators, as stated by BCCIM) 
one finally
derives the range of luminosity  given in Table~\ref{tab7} under the two
alternative assumptions about the mode of pulsation. One is 
now in the position of comparing observed lightcurves with the
atlas presented by BCCIM and BCCM. For the various stars one finds:

\medskip

{\bf CN Lyr:} If F pulsator, would be out of the range explored by theory
when Z=0.01. However, comparison also with results for Z=0.02
suggests that the theoretical rising time of a F pulsator should
be much shorter than observed. Comparison again with theoretical
predictions for Z=0.01 and Z=0.02 could suggest that this star
should be a FO pulsator crossing the strip at large luminosity well 
above the ZAHB. Even though the star is  beyond the limit of the atlas, 
one finds, e.g., that Z=0.02 FO pulsators with similar temperatures and large
luminosities ($\rm \log{{L}/{L_{\odot}}}$= 1.61 and 1.81) show the asymmetric 
lightcurve disclosed by our observations. Note that in this case
one cannot derive intrinsic color from the quoted Caputo \& De Santis (1992) 
relation, and the reasonable prediction about the CN Lyr reddening should
be regarded as obtained by chance. However, as suggested by our
referee, on observational ground CN Lyr (like, e.g., FW Lup in Lub 
(1977) sample) has to be regarded
as a typical low-amplitude, type b fundamental pulsator, as found
towards the red edge of the instability strip. Appropriate theoretical
investigation are needed before discussing possible mismatches with theory.

\medskip

{\bf AQ Lyr:} There is, apparently, no way to fit the observed lightcurve 
to theoretical predictions for FO pulsators. The shape
of the curve is in good agreement with predictions for F pulsators
in the quoted range of luminosities and temperature. Note that 
the luminosity is the one predicted by BCCIM for metal rich
HB stars. However, bearing 
in mind that bolometric amplitude are roughly comparable with
amplitude in the V band (Marconi, private communication) theory predicts larger amplitude.
A similar occurrence has been already discussed in BCCM (see
Fig. 18 in that paper). 

\medskip

{\bf AW Dra:} It appears a classical ab type pulsator. As a matter
of the fact, in BCCM one finds that a metal poor FO pulsator 
at the required large luminosity should have a symmetric lightcurve,
contrarily to observation. Both amplitude and shape of the lightcurve 
appear in reasonable agreement with predictions for F pulsators,
even if the atlas lack models in the range $\rm T_e = 6500-6800K$. Thus 
one derives that AW Dra appears as a fundamental pulsator crossing
the strip at $\rm \log{{L}/{L_{\odot}}} \sim  1.8- 1.9$, i.e. above the ZAHB 
luminosity level.

\section{Conclusions}

In this paper we presented lightcurves for the three field 
RR Lyrae AQ Lyr, CN Lyr and AW Dra. As for the two metal rich
pulsators in the sample we suggest that CN Lyr has been possibly
misclassified, being possibly an asymmetric c-type pulsator.
This gives a warning against the possible contamination with
FO pulsators of observational samples of field ab type stars.
However, CN Lyr could also be a b-type fundamental pulsator
with lightcurve not well predicted by the available theory.
AQ Lyrae appears as a metal rich ab type pulsator, with the 
expected luminosity but with amplitude lower than predicted by theory.
Finally, we find that AW Dra  behaves as a metal poor fundamental 
pulsator, crossing
the instability strip well above the corresponding ZAHB 
luminosity level.

\begin{acknowledgements}
It is a pleasure to thank Dr. M. Marconi for providing the data of 
Fig.~\ref{fig4} in computer form and for many helpful 
suggestions and discussions. 
We also wish to thank Dr. A. Alonso, Dr. S. Gentili, and Dr. F. D'alessio for  
their collaboration during the observations.  
Special thanks go to Dr. G. Bono and Dr. E. Brocato for many useful 
discussions and to our referee, Jan Lub, for carefully
reviewing the paper and for helpful suggestions.
\end{acknowledgements}

{}

\end{document}